# A Vehicle Size-Based Dynamic Model of Artificial Driving Risk Potential Fields and Vehicle Interaction Analysis for Highway Driving

Ru Ling, Meng Li, Bowen Liu, and Zhibin Li

*Abstract*—**Traditional driving risk potential field model generally assumes uniform vehicle sizes, which fails to reflect the heterogeneity in real-world traffic environment. This study aims to develop an improved model by introducing the vehicle dimensions into the driving risk potential field framework. Vehicles are categorized based on size into standard-sized vehicles, medium-sized vehicles, and large-sized vehicles. And the interactions between two vehicles and the force field excited by the driving risk potential field are investigated in this paper. Our analysis of the risk field contours at various vehicle speeds and sizes reveals that larger and faster vehicles generate more intense and expansive risk fields, necessitating increased following distances. For larger vehicles, the risk field extends into adjacent lanes. Force field analysis demonstrates that vehicles traveling in adjacent lanes undergo changes in their motion state due to the influence of the force field emanating from the neighboring larger vehicles. The model effectively points out that, for adjacent lanes, the regions surrounding the tail and head of high-speed vehicles represent highly risky zones.**

*Index Terms*—**Risk Potential Field, Vehicle Dimensions, Force Field, Highway driving.**

## I. INTRODUCTION

FOR decades, researchers have extensively explored the complex dynamic characteristics of traffic flow, which is a fundamental area with wide-ranging practical applications in traffic engineering, self-driving vehicle development, and transportation safety. In recent years, advancements in sensing technology and the widespread adoption of big data have made it possible to obtain high-precision vehicle trajectory data, renew the exciting interest in data-driven modeling methods in this field. However, it is essential to recognize that analytical models based on the principles of physics and control theory play a key role in explaining the underlying fundamental laws governing traffic flow phenomena. These physics-based models have stronger understandability and also help researchers and professionals make more informed decisions to improve the efficiency and safety of transportation systems.

Ru Ling, Bowen Liu and Zhibin Li are with the Department of Transportation, Southeast University, Southeast University Road #2, Nanjing, China, 211189 (e-mail: ruling@seu.edu.cn, sivanliu@seu.edu.cn, lizhibin@seu.edu.cn).

Meng Li is with School of Mechanical and Aerospace, Engineering, Nanyang Technological University, Singapore 639798 (e-mail: mengli@ntu.edu.sg).

As a typical example of an analytical model, the application of artificial potential field theory in the automotive industry has become widespread in describing traffic risks. This theory requires the use of a risk potential field, commonly used for robot path planning. In this model, the robot moves towards its destination through the attractive field generated by the target itself, while obstacles are avoided through the repulsive field surrounding them. In 1999, Gerdes and Rossetter [1] introduced this approach into the transportation domain, where they developed a driver assistance system based on the risk potential field concept. They used potential fields to mark dangerous regions between lanes and created a control law to ensure vehicles stay within their designated lane according to the field intensity. Subsequently, Wolf et al. [2] described the various components of the risk field's potential function for modeling highway driving. These components include lane potentials, collision avoidance vehicle potentials, and velocity potentials. Furthermore, a pioneering concept known as the "driving safety field" [3] uses field theory to describe risk factors from drivers, vehicles, and road conditions. Based on this, Li [4] introduced the key parameter of acceleration into the driving risk field, enhancing the risk potential field's sensitivity to the vehicle's motion state. Subsequently, Ref. [5] and [6] explored drivers' reactions within the driving risk potential field to comprehensively assess the impact of drivers in potential traffic accidents. Furthermore, article Ref. [7] applied non-cooperative game theory to the driving risk potential field model, aiming to further optimize and enhance driving strategies.

The potential field method presents significant advantages, particularly in its simplicity and efficiency. Even within complex road environments, it maintains low computational costs and can generate smooth vehicle trajectories in real time. Consequently, the potential field method has quickly gained attention as a prominent research area, covering obstacle avoidance [8,9,10], car-following models [4,11], and lane-changing models [12,13,14]. Artificial potential fields are crucial components in the realm of path prediction and planning [15,16,17,18], providing theoretical and algorithmic frameworks for higher-level autonomous driving and path optimization. Moreover, in the field of traffic flow, the theory of risk potential fields has made significant progress, especially in avoiding collisions, leading to the development of several optimized model algorithms [19,20,21].

Although the current driving risk potential field models effectively assess the risk levels in driving environments, the development of theoretical research still faces significant challenges. On one hand, current models of driving risk potential fields lack effective means to describe the interactions between vehicles. It is well-known that complex, variable, and uncertain traffic scenarios arise from the interactions and reactions among high-speed vehicles. Without tools to describe these interactions, understanding and analyzing various traffic behaviors and phenomena presents substantial challenges. On the other hand, to avoid excessive idealization, theoretical research should focus on using the safety potential field model to simulate more realistic traffic scenarios as closely as possible. This idealization is particularly evident in the common assumption of uniform vehicle sizes in traditional driving risk potential field models. Therefore, a critical task for the further development of driving risk potential field models is to incorporate vehicle sizes into the current model framework and to seriously consider how to examine the interactions between vehicles of different sizes.

It's important to emphasize that vehicle size is a significant factor affecting driving risk, yet it has not been given adequate attention in past research. The reason behind this is that larger vehicles have larger blind spots, greater force, and more complex handling abilities, greatly reducing their ability to avoid accidents. Additionally, the aerodynamic structure around large vehicles tend to be less stable when traveling at high speeds. The severity of accidents caused by vehicles of different sizes varies greatly, directly related to the degree of injury and economic loss. These objective factors highlight the increased risks associated with driving larger vehicles on highways. This paper aims to delve into the characteristics of highway risk potential field models from a theoretical perspective, with a particular focus on the influence of vehicle size.

Another focus of this paper is the examination of interactions between vehicles based on the driving risk potential field. From the perspective of field theory, it is quite natural to analyze the logic of vehicle interactions through the interactions of fields. In other words, the potential field spontaneously gives rise to a corresponding force field. The dynamic characteristics of vehicles within this force field represent the modes of interaction between vehicles, which have not been effectively analyzed or discussed in the current driving risk potential field models.

Therefore, this paper introduces an improved driving risk potential field model for highways that considers vehicle size factors, exploring the interactions between large and small vehicles due to traffic risks. This model integrates vehicle size into a parametric mathematical framework, thereby extending the existing theory. Based on the dimensions of vehicles and the current methods of traffic data collection, vehicles are classified into standard-sized, medium-sized, and large-sized categories. Through modeling of the driving risk potential field, we examined the spatial distribution of risk potential fields formed by vehicles of varying speeds and sizes. Additionally, by analyzing the force field, we investigated the interactions between large-sized vehicles and standard-sized vehicles, particularly the changes and patterns in the motion states of standard-sized vehicles traveling near large-sized vehicles, and the corresponding risk assessments.

The remaining sections of the paper are structured as follows: Section II provides a detailed overview of the model construction process. Section III demonstrates the application of the model. Section IV delves into the dynamic analysis of interactions between vehicles. Section V focuses on data statistics and validation. Conclusions and discussions are presented in Section VI.

## II. MODEL METHODOLOGY AND CONSTRCUTION

In this section, our focus centers on elucidating the construction of the potential field model for vehicle behavior on highways. The highway traffic environment primarily comprises two components: the dynamic movement of vehicles and the static elements. Firstly, the presence of large vehicles on the road significantly escalates driving risks. Thus, in this model, we examine how the size of vehicles influences the spatial distribution of potential fields and their interactions. Secondly, the static environment encompasses lane markings, signs, and road boundaries. It's noteworthy that stationary vehicles with zero speed also constitute static elements in traffic, including obstacles on the road. These environmental factors contribute to irregular or more complex road boundaries.

Therefore, according to Ref. [2], the risk potential field can be divided into three primary components: the lane potential field, the road potential field, and the vehicle potential field. The lane potential field and road potential field signify static elements in traffic, while the vehicle potential field encapsulates dynamic vehicle interactions within this static environment. For the sake of model simplification, we assume the absence of obstacles or stationary vehicles with zero speed on the highway.

Before commencing the modeling process, let's establish our reference frame as follows: the longitudinal motion of vehicles aligns with the y-axis, while the lateral direction corresponds to the x-axis. As shown in Fig. 1, vehicles progress in increasing x-directions, commencing from position $x = 0$ for ease of reference, with $y = 0$ denoting the boundary line of the bottom lane. We posit the existence of $N$ lanes, each with a width of $D$, and vehicles traverse along an infinitely long straight roadway. It's pertinent to acknowledge that while vehicles may navigate curved lanes, their spatial distribution continues to mirror that of straight roads. However, delving into the implications of curved lanes on the interactions between vehicle potential fields and lane/road potential fields extends beyond the scope of this paper.

The details regarding vehicle dimensions are as follows: The vehicle width is denoted by $Y$ which satisfies $Y < D$; the vehicle length is represented by $X$, and the vehicle height is denoted by $W$. The influence of vehicle dimensions on risk potential fields will be intuitively elucidated in the subsequent section dedicated to the vehicle potential field.

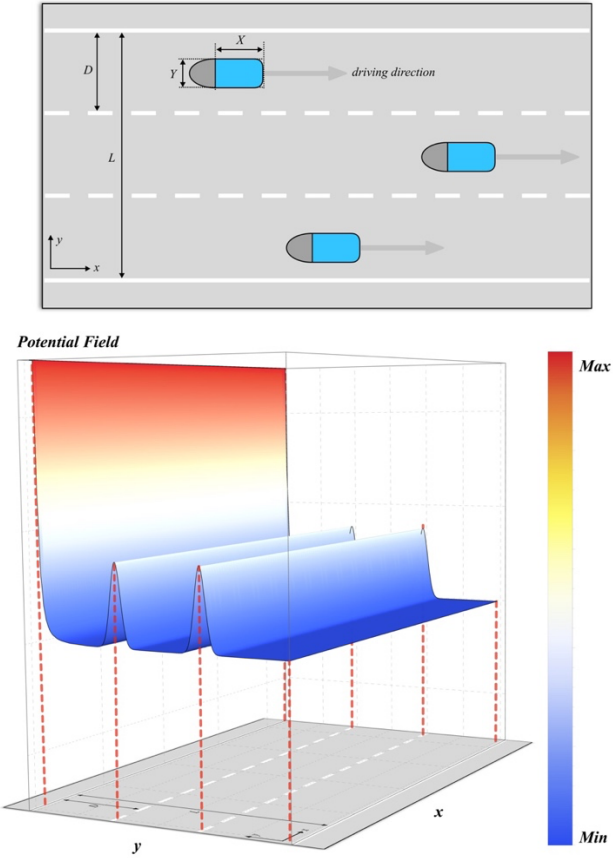

**Fig. 1.** Highway coordinate frames and potential field of load and lanes.

## A. Road Potential Field

The road potential field primarily delineates the boundaries within which vehicles travel on highways. Given that vehicles are constrained to maneuver within designated lanes, the boundaries on both sides create an infinitely deep potential well. This well features infinitely high walls along the boundaries on either side of the road, gradually diminishing towards the center of the road. Consequently, vehicles are constrained to operate within this potential well. The description of this potential well is captured by the following equation

$$U_{road} = \frac{1}{2} A_{road} \left( \frac{1}{y^2} + \frac{1}{(y-L)^2} \right), \quad (1)$$

here $A_{road}$ serves as a scaling factor, with its magnitude typically dictating the steepness of the potential barrier near the boundary. Additionally, $L = N \times D$ is defined as the width of a one-way road, equating to the lane width $D$ multiplied by the number of lanes $N$. It's imperative to note that our current consideration involves the simplest case, disregarding scenarios involving lane merging or splitting at specific locations on the road. While such cases would render the problem more complex, they would also render the model more applicable to real-world scenarios.

## B. Lane Potential Field

On highways, vehicles adhere to designated lanes for travel. To ensure safe navigation within a lane, height barriers are often installed at the lane markings, guiding vehicles effectively. Consequently, when a vehicle intends to switch lanes, it must elevate its "kinetic energy" beyond the height of the lane barrier. In this context, "kinetic energy" encompasses strategic factors influencing the driver's inclination to change lanes, such as encountering slow-moving vehicles ahead in the lane or adjusting speed levels approaching the maximum or minimum speed limit of that lane.

Depending on actual conditions, the height of barriers on lane markings must strike a balance: they need to be sufficiently high to facilitate vehicles traveling along the center of their respective lanes, yet low enough to allow for lane changes based on prevailing lane conditions. Hence, the form of the lane potential field can be expressed as

$$U_{lane} = \sum_{i=1}^{N-1} A_{lane} \, e^{-\frac{(y-i\times D)^2}{2\sigma^2}}, \quad (2)$$

where $A_{lane}$ determines the height of the barrier on the lane line, while $\sigma$, as a scaling factor, describes the steepness of the barrier on the lane line. The example of space distribution of the lane and load potential field can be referred in Fig. 1.

## C. Vehicle Potential Field

The vehicle potential field, often referred to as the risk potential field, originates from vehicles traversing highways. As an observer maneuvers their vehicle in proximity to other target vehicles, they encounter the influence of the potential field generated by these vehicles. Therefore, the essence of modeling highway potential fields hinges on incorporating real-time dynamic constraints and the impacts of vehicle potential fields, which are shaped by all surrounding vehicles, on the movement of the observer's own vehicle.

*a). Space distribution:* According to Ref. [2], we design vehicle potential fields using Yukawa potentials:

$$U_{vehicle} = \sum_i A_{vehicle}^i \frac{e^{-\alpha_i K_i}}{K_i}, \quad (3)$$

where $A_{vehicle}^i$ denotes the magnitude of the vehicle potential field for the $i$th target vehicle. It is widely recognized that the magnitude of a potential field determines the level of risk, which correlates closely with the speed of target vehicles. Assuming the speed limit for this lane is $v_m$, as a target vehicle's speed approaches this value, its associated risk escalates. Conversely, if a vehicle surpasses this speed limit, we can artificially induce a sharp increase in its risk potential field. Hence, without loss of generality, let us assume that

$$A_{vehicle}^i = A_m^i \, e^{\frac{v_i - v_m}{v_m}}, \quad (4)$$

$A_m^i$ represents the normalized scaling factor of the $i$th target vehicle, and the magnitude of the vehicle potential field escalates rapidly with the increasing speed of the target vehicle. It is essential to highlight that we have not addressed the scenario where the vehicle speed is exceptionally low here. Clearly, driving at a very low speed on a highway poses significant risks as well. We presume that vehicles on highways uphold a relatively high and consistent speed without surpassing lane speed limits.

Moreover, $K_i$ represents the effective distance in the vehicle potential field, illustrating how the vehicle potential field diminishes in space. Thus, it serves as a pivotal factor in establishing the vehicle potential field and is intricately linked to various driving factors. Primarily, the driving speed of vehicles is a critical determinant. It's well-established that increasing the speed of the target vehicle elevates driving risks. This not only impacts the magnitude of the vehicle potential field but also amplifies the overall risk perception within the vicinity of the target vehicle. Consequently, the zone where traffic accidents are likely to occur expands correspondingly. Hence, the effective distance is regulated by vehicle speed. It's imperative to underscore that our refinement of previous risk potential fields hinges on considering the mutual interactions between vehicles' potential fields. As a result, we define our scale factor as

$$\xi_i = \xi_{0i}(v_i) \prod_j e^{-\beta(v_i - v_j)}, \quad (5)$$

where $v_j$ describes the driving speed of other vehicles adjacent to the target vehicle $v_i$ on the same lane, including both other target vehicles and controlled ones. $\xi_{0i}(v_i)$ serves as a normalization factor and is also influenced by vehicle speed. In this context, we reference Ref. [2], where they employ "the three-second rule," signifying that following vehicles must sustain a distance equivalent to three seconds of travel time behind preceding cars

$$\xi_{0i}(v_i) = \prod_j \frac{d_0}{T_f v_i}, \quad (6)$$

as $T_f = 3$ s, $d_0$ represents the influence distance of the vehicle's potential field in the radial direction.

*b). Vehicle dimensions:* More importantly, we will now consider the influence of vehicle size on the vehicle's potential field. Firstly, when the vehicle size increases, such as the presence of large buses or trucks, it significantly heightens the risk for vehicles traversing the road. A particularly intuitive observation is that vehicle size directly influences the magnitude of the vehicle's potential field. Therefore, we must consider

$$A_m^i = A_m \frac{X_i \, Y_i \, W_i}{X_0 \, Y_0 \, W_0}, \quad (7)$$

as $A_m$ is a unified normalization scale factor, $X_i, Y_i, W_i$ are respectively the length, width and height of $i$th target vehicle; while $X_0, Y_0, W_0$ are respectively the length, width and height of standard-sized vehicles on highways. We define the size of the smallest vehicles on highways as the standard size for our model, based on the dimensions of compact cars. Consequently, as vehicle sizes increase, the magnitude of their potential fields also increases accordingly.

Secondly, vehicle size also influences the effective distance. The criterion we adopt for quantification is straightforward: for surrounding vehicles beside target vehicles, the increased length and height would expand blind spots in drivers' vision and intensify driving difficulty, thereby heightening road risks.

Similarly, for front or rear vehicles relative to target ones, increased width and height would also elevate road risks. Hence, we comprehensively define the effective distance as

$$K_i = \sqrt{\left((x - x_0)\xi_i\right)^2 \left(\frac{Y_0 \, W_0}{Y_i \, W_i}\right)^{\frac{1}{3}} + (y - y_0)^2 \left(\frac{X_0 \, W_0}{X_i \, W_i}\right)^{\frac{1}{3}}}, \quad (8)$$

where $(x_0, y_0)$ describe the shape of the target car.

After incorporating the vehicle dimensions into the model, it's crucial to underscore that the geometric shape of the vehicle also significantly impacts the distribution of its potential field. Drawing from Ref. [2], we suggest introducing a curved area at the rear of the vehicle, as depicted in Fig. 1. We opt for a curved area over a wedge-shaped one due to our consideration of the lane-changing effect in this model. Wedge-shaped areas with sharp corners tend to generate overly sharp potential fields. In real driving scenarios, as a controlled vehicle approaches a target vehicle in front, there is indeed a propensity to change lanes. Nevertheless, we maintain the belief that, at this juncture, the potential field generated by the target vehicle should be smoother and more rounded. Particularly on highways with dense traffic flow, we must acknowledge that the lane-changing tendency of controlled vehicles may not be as urgent or abrupt.

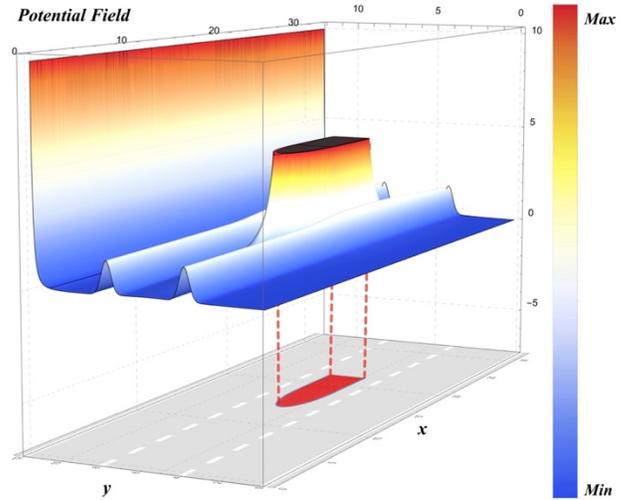

**Fig. 2.** Spatial distribution of the total risk potential field.

It is worth further emphasizing that on highways, the lane-changing behavior of controlled vehicles primarily relies on drivers' subjective judgment. While the proximity of approaching vehicles does influence drivers' decision-making processes, it is not their primary consideration. Therefore, the potential field formed by target vehicles ahead serves as one influencing factor for drivers' lane-change decisions, and our intention was for it to exert only a mild impact on controlled vehicles behind them. This primary intention underscores the introduction of curved regions, which demonstrate excellent adaptability for subsequent model simulations, albeit increasing the modeling difficulty accordingly.

### D. Potential Field Distribution

The total risk potential fields on the highway result from the combined effects of various environmental factors, including the road potential field, lane potential field, and vehicle potential field

$$U_{all} = U_{road} + U_{lane} + U_{vehicle}. \quad (9)$$

As a simple example, Fig. 2 illustrates the total risk field formed around a car traveling at a speed of $v = 20$ m/s on the highway. The curved area at the rear of the vehicle serves as an artificially extended zone that influences the driving behavior of following vehicles and visually extends the risk potential field formed by vehicles. The height of the potential field signifies the level of risk, allowing us to discern how the risk potential fields are distributed on the highway. The primary contributions at $y = 0$ and $y = 12$ m stem from road potential fields, while those at $y = 4$ m and $y = 8$ m reflect lane potential fields. This distinction is more apparent in Fig. 3 from left view.

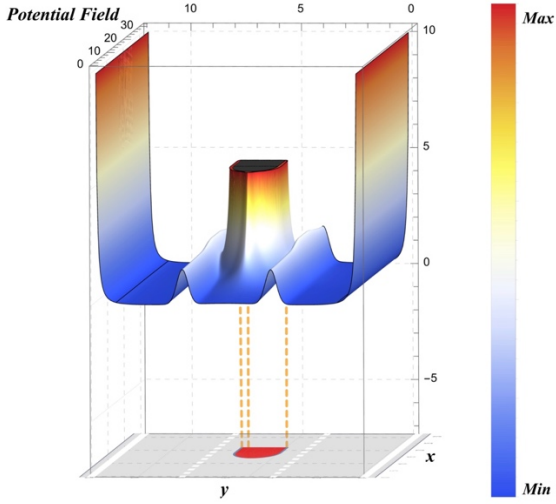

**Fig. 3.** Left view of the total risk potential field.

Furthermore, in this left view, we observe an expansion of risk potentials near lanes where vehicles are present. This expansion arises from the combined effects of lane potentials and vehicle potentials in this area. Thus, we can discern interactions and influences between different types of potentials. A more detailed discussion on this aspect will enhance the modeling and practical applications related to vehicle risk potentials. From this figure, it becomes evident that when a controlled vehicle closely approaches another vehicle, its risk potential sharply increases; maintaining too short a distance between two vehicles significantly escalates driving risks.

## III. MODEL APPLICATION

In this section, we will delve into the impact of various parameters on the total risk potential field, focusing particularly on the influence of vehicle speed and vehicle dimensions on the spatial distribution of risk potential fields. Before embarking on model simulation, we first establish and define some of the model parameters. Without loss of generality, in this model, we assume the lane width $D = 4$ m and lane count $N = 3$, resulting in a one-way road width $L = 12$ m. For standard vehicle dimensions, we define length $X_0 = 4$ m, width $Y_0 = 2$ m, and height $W_0 = 1.5$ m. Please refer to Tab. 1 for detailed parameter settings.

TABLE I
MODEL PARAMETER SETUP

| | | | |
|---|---|---|---|
| Highway Setup | $D$ | Lane width | 4 m |
| | $N$ | Lane number | 3 |
| | $L$ | Road width | 12 m |
| | $v_m$ | Speed limit for the road | 30 m/s |
| | $X_0$ | Standard vehicle length | 4 m |
| | $Y_0$ | Standard vehicle width | 2 m |
| | $W_0$ | Standard vehicle height | 1.5 m |
| Road Potential Field | $A_{road}$ | Scaling factor | 1 |
| Lane Potential Field | $A_{lane}$ | Field amplitude | 1.5 |
| | $\sigma$ | Width scaling factor | 1/4 |
| Vehicle Potential Field | $A_m$ | Unified normalization scale factor | 6 |
| | $\alpha_i$ | Yukawa scale | 1/2 |
| | $T_f$ | Following time | 3 s |
| | $d_0$ | Influence distance | 10 m |

### A. The impact of vehicle speed on the risk potential field

In this subsection, we will explore the impact of vehicle speed on the total risk potential field. The faster a vehicle travels on a highway, the higher the likelihood of traffic accidents occurring, and the more uncontrollable the consequences become. For instance, if a controlled vehicle is following closely behind another target vehicle that suddenly decelerates at high speed, the risk of collision between them significantly increases. Similarly, if a controlled vehicle is ahead of another target vehicle that is speeding or decelerates abruptly, there is also a heightened risk of collision from behind. These risks are prevalent on real highways, and our model can quantify the level of risk posed by vehicles traveling at high speeds by modeling their potential fields.

Here, we assume that standard vehicles are driving on the highway with car dimensions, i.e., $X = X_0, Y = Y_0$ and $W = W_0$. In Fig. 4, we illustrate how the distribution of the total risk potential field varies in space when target vehicles are traveling at speeds of 10 m/s, 20 m/s, and 30 m/s. It is evident that as speed increases, the risk potential field around vehicles rapidly spreads throughout space. The lane behind a vehicle is particularly impacted by this spread; as speed escalates further, the risk potential field expands to even greater distances behind vehicles. This implies that maintaining a safe distance between two vehicles within the same lane increases as speed rises. Therefore, controlling for longer following distances becomes imperative to mitigate driving risks for controlled vehicles. This conclusion aligns with regulations governing safe driving practices and underscores the effectiveness of our model.

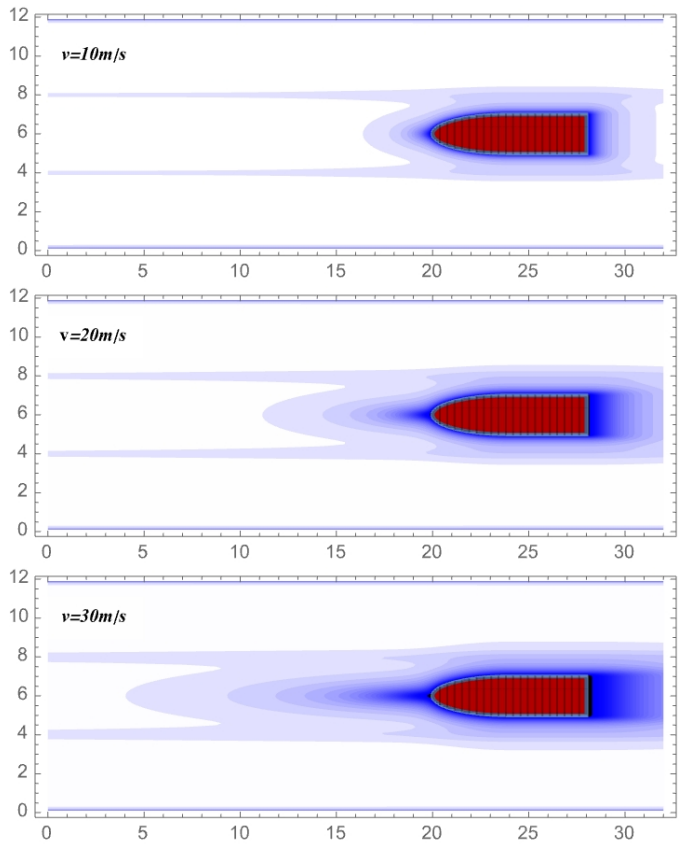

**Fig. 4.** Contour plots of the total risk potential field at different vehicle speed.

Simultaneously, we also observe that as vehicle speed increases, there is a concurrent rise in the risk potential field ahead of it. This aspect distinguishes our model from previous ones. Here, we do not suppress changes in the risk potential field ahead of vehicles because objectively, this area in front of high-speed moving vehicles constitutes a high-risk zone. A typical scenario arises when a side-vehicle intends to change lanes into another lane occupied by our target vehicle; they must carefully gauge their distance from our controlled vehicle. If they are too close in proximity, it could lead to failed lane change attempts or even serious accidents, underscoring the presence of driving risks within this forward area for vehicles as well. Once more, our vehicle model quantitatively illustrates previously overlooked aspects regarding vehicular behavior and driving risks mentioned in earlier articles.

### B. The impact of vehicle dimensions on the risk potential field

Next, we will analyze the impact of vehicle dimensions on the distribution of risk potential fields. This is a crucial factor that has received relatively little discussion in previous articles. As we all know, there is a substantial variance in vehicle sizes on highways and in other driving scenarios. Vehicles of different sizes occupy differing amounts of space, which is one aspect influencing the distribution of risk potential fields for vehicles and has been considered in many modeling studies. However, varying vehicle sizes can lead to significant differences in blind spots, driving inertia, and maneuverability, resulting in noticeable variations in the ability to avoid accidents during the driving process. Moreover, vehicles of different sizes also cause vastly different levels of accident severity, which directly correlates with the degree of injuries and economic losses. This results in significant variations in their respective field distributions. This underscores one of the main reasons why we introduce factors such as body size.

Firstly, for ease of discussion, we simplify the vehicle model and divide them into three categories based on body size. The first category consists of standard-sized vehicles mainly including small cars, station wagons, SUVs and other compact vehicles. The second category comprises medium-sized vehicles such as mid-size buses or motorhomes with defined dimensions: length $X = 8\,\mathrm{m}$, width $Y = 2.5\,\mathrm{m}$, and height $W = 3\,\mathrm{m}$. Lastly, the third category encompasses large-sized vehicles including large buses, trucks, or trailers with defined dimensions: length $X = 12\,\mathrm{m}$, width $Y = 3\,\mathrm{m}$, and height $W = 4\,\mathrm{m}$. We then consider these three types as examples traveling at a speed of $10\,\mathrm{m/s}$ on highways to analyze their total risk potential field formations.

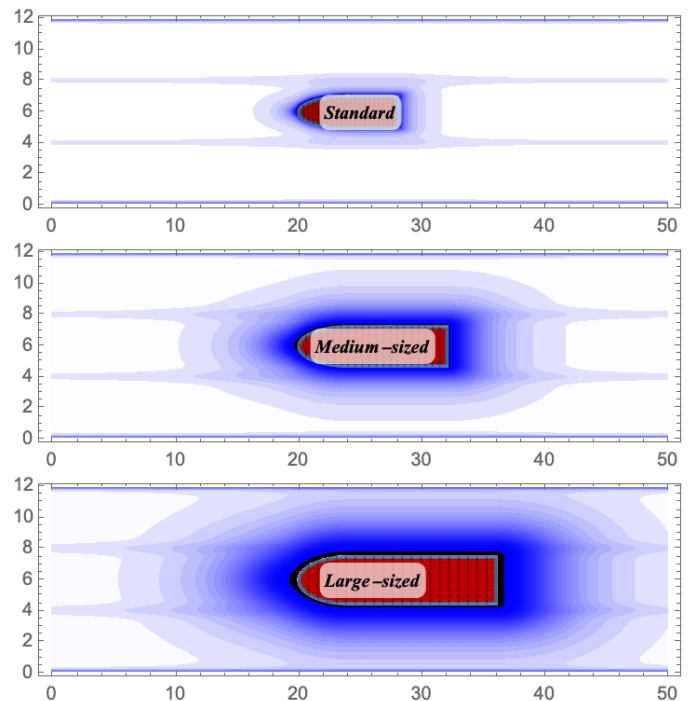

**Fig. 5.** Contour plots of the total risk potential field at different vehicle speed.

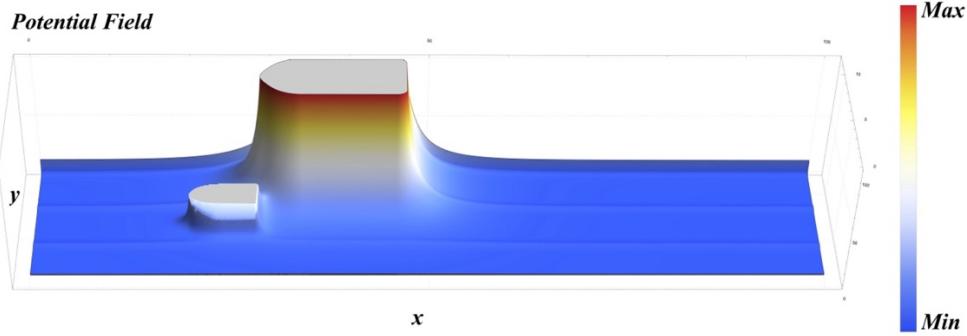

**Fig. 6.** The total risk potential field generated by the two vehicles traveling at a speed of 10 m/s inside the road.

Conversely, small cars exhibit relatively weak risk potential fields formed on their sides and hardly cross lanes to affect the safety of vehicles in adjacent lanes. However, with medium-sized vehicles, we can visually observe that an increase in vehicle length and height exerts certain driving pressure on side vehicles, causing the risk potential field to extend into adjacent lanes. Even in the case of large-sized vehicles, there is a considerable degree of diffusion of their risk potential fields into adjacent lanes, thereby increasing driving risks for vehicles in those lanes.

These findings corroborate our driving experience, indicating that larger-sized vehicles pose significant risks on the road due to their increased dimensions in length, width, and height. These vehicles entail large blind spots and exhibit more complex maneuverability, as well as a more intricate airflow structure around them. Consequently, driving alongside or behind them becomes more perilous. Moreover, the presence of larger-sized vehicles significantly impairs visibility for drivers in controlled vehicles. To mitigate these risks, it is advisable for larger-sized vehicles to travel along the edge lane (e.g., bottom lane) to reduce their impact area within the vehicle's potential field and minimize their influence on traffic in distant lanes. This measure ensures safer driving on highways.

## IV. Dynamic Analysis of Interaction

The vehicle risk potential field model constructed in this paper facilitates intuitive and clear observation of driving risks around the target vehicle by calculating the spatial distribution of the risk potential field. However, the key point lies in the model's ability to naturally incorporate interactions between high-speed vehicles. Examining these interactions not only enhances the authenticity and efficiency of model simulation but, more importantly, aids in analyzing real traffic behaviors.

The driving risk potential field serves as a comprehensive physical model, conceptualized as a function of potential field. Within this framework, objects demonstrate diverse potential energy states across spatial coordinates, influencing their motion trajectories and behavioral patterns. A fundamental aspect of the potential field is its scalar nature, meaning that each spatial point is characterized by a potential energy value. Consequently, the driving risk potential field illustrates the distribution of risks across spatial positions, allowing for a profound understanding of the juxtaposition of high-risk areas and relatively safer areas.

However, to elucidate the interactions between vehicles, it becomes imperative to explore the variations in potential energy across different spatial directions at specific points. From the perspective of field theory, this entails computing the spatial configuration of the force field generated by the interaction-modified potential field. It's essential to recognize that this force field emerges directly from the potential field, enabling not only numerical assessment of risk distribution but also intuitive comprehension of the driving risk forces influencing controlled vehicles in space. Thus, the simulation of highway traffic behaviors becomes dynamically feasible. To accomplish this, the force field of the driving risk potential field is defined as its gradient

$$\boldsymbol{F} = -\nabla U_{all}(x, y) \qquad (10)$$

with the symbol $\nabla$ representing the gradient operator as $\nabla = \vec{i}\,\partial/\partial x + \vec{j}\,\partial/\partial y$. Hence, by computing the force field, we can discern the driving risk forces exerted on the controlled vehicle as it approaches the target vehicles, stemming from the interaction among risk potential fields. Analyzing the magnitude and direction of these forces enables us to precisely comprehend the dynamic state of the controlled vehicle, thereby facilitating further research into its traffic behavior.

Without loss of generality, we examine the interaction between two vehicles as a straightforward example in the ensuing discussion. Our attention is directed towards the interaction between vehicles to avoid interference from static environments and to reduce our computational burden. Therefore, we only calculate the force field distribution formed by the interaction-modified vehicle potential field and analyze its outcomes. As an example of this interaction case in our model, we plot the 3-dimensional distribution of the risk potential field in Fig. 6, which clearly highlights the difficulty of capturing the interaction between two vehicles solely from the potential field.

Fig. 7 illustrates the force field distribution experienced by any point in the surrounding space as a standard-sized vehicle, traveling at a speed of 10 m/s on a highway, encounters a large-sized vehicle moving at the same speed. We have plotted the large-sized vehicle and related force field simultaneously under identical conditions to contrast their disparities. while the color intensity, which corresponds to temperature variation, signifies the magnitude of the force.

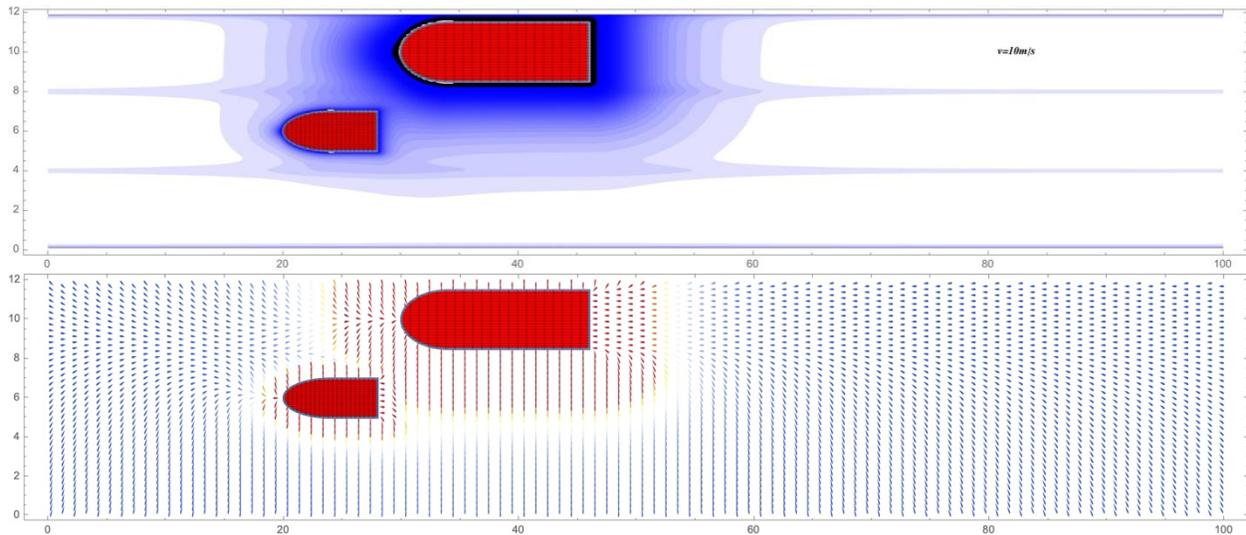

**Fig. 7.** A scenario depicting the imminent encounter between a standard-sized vehicle and a large-sized vehicle, each traveling in separate lanes. Above: The total risk potential field generated by the two vehicles traveling at a speed of 10 m/s in different lanes. Below: The vector field of forces generated by the two vehicles at the same speed.

As shown in the Fig. 7, the blue droplet-shaped arrows represent the weak force field formed by the vehicle's risk potential field, while the yellow and red areas depict regions with intense force field distribution. This allows us to clearly discern the force field distribution enveloping the vehicle, where the force field area of large-sized vehicle is significantly larger than that of standard-sized vehicle. Furthermore, the force field at the rear of the vehicle radiates and diffuses towards the area behind the vehicle from the midpoint of the arc, while the force field at the front of the vehicle propagates forward along the perpendicular line from the vehicle's front end.

It is noteworthy that when standard-sized vehicles approach large-sized vehicles, the strong interaction between their risk potential fields causes the force field at the front of the standard-sized vehicle to lean entirely towards the bottom lane. This implies that the controlled vehicle (standard-sized vehicle) continuing to travel experiences a tendency to change lanes away from the large-sized vehicle due to the downward force generated by the presence of the large-sized vehicle. In this scenario, if the controlled vehicle does not change lanes, it will continue to be influenced by the force field generated by the large-sized vehicle's risk potential field, thereby altering its dynamic characteristics. This indicates that the force field produced by large-sized vehicles extends, leading to intense interactions with vehicles in adjacent lanes.

It is important to note that while the distribution of risk potential fields can clearly indicate the spatial distribution of driving risk, analyzing the interaction between driving vehicles necessitates an objective assessment of the distribution of driving risk force fields. This need becomes increasingly evident as vehicle speeds rise. For example, in Fig. 8, we examine the spatial distribution of risk potential fields and risk force fields when the vehicle speed is 25 m/s. If we consider that in Fig. 7, the controlled vehicle only partially entered the edge of the target vehicle's risk force field, then in Fig. 8, we observe that the controlled vehicle has penetrated the occupied area of the target vehicle's risk force field to a considerable extent, resulting in much more intense interactions compared to Fig. 7. This heightened interaction is reflected in the extent to which the droplet-shaped arrows representing the force field at the front of the controlled vehicle tilt towards the lower lane, indicating deeper interactions. Additionally, the disorganized droplet-shaped arrows in the side and rear areas of the controlled vehicle suggest intense dynamic interactions.

Furthermore, the spatial distribution of the risk potential field highlights a significant increase in the risk area surrounding large-sized vehicles as vehicle speed rises. This escalation in risk is particularly noticeable within the current lane, where the risk potential fields of the target vehicle's front and rear extend noticeably farther. However, concerning the risk force field, this increase in risk is not uniformly linear. Firstly, in the rear part of the vehicle, the force field distribution exhibits a swallowtail shape. This occurs because the risk potential field at the rear steepens with increasing speed, causing the force field on both sides of the lane to be much larger than those at the midpoint of the arc. Consequently, vehicle following behind the target vehicle displays a stronger inclination to change lanes toward the adjacent lanes on both sides. On the other hand, the risk force field in front of the vehicle intensifies in both intensity and coverage area with speed, while the force field in the side front space of the vehicle becomes significantly more intense and expansive. Theoretically, this situation arises from the vector addition property of driving risk force. Thus, it implies that there is a considerable driving risk in the side front and rear areas of the target vehicle.

Combining the analysis of the rear force field, we can infer that although the vehicle following behind a large-sized vehicle has a strong inclination to change lanes due to the presence of force field, the driving risk after changing lanes does not decrease effectively. From a strategic perspective, it is safer to reduce speed until away from the large-sized vehicle before changing lanes (while continuous lane changes may significantly reduce risks, considering vehicles traveling at high speeds in other lanes, this is not a safe and prudent strategy).

Furthermore, in terms of the risk potential field, the influence of large-sized vehicle extends to the adjacent third lane. However, regarding the risk force field, this impact on vehicles traveling in the third lane is constrained. Vehicles in adjacent lanes will persistently experience the force field generated by the risk potential field of neighboring large-sized vehicle as they pass it. Analyzing the dynamic characteristics of these vehicles can help confirm the validity of our model. In the following section, we will validate and analyze the effectiveness of our model based on this observation.

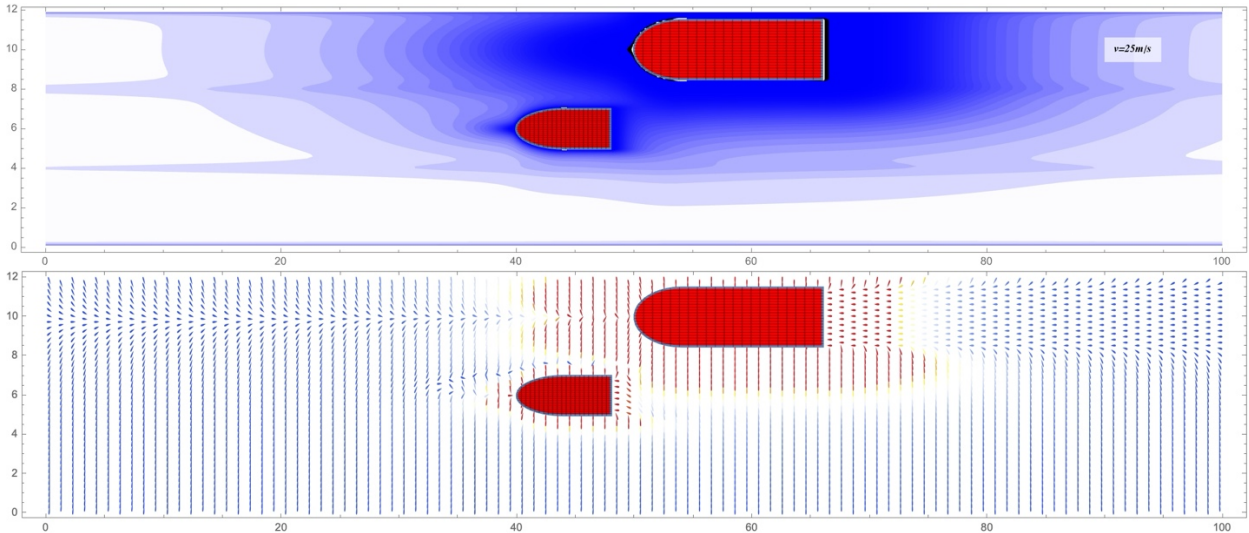

**Fig. 8.** A scenario illustrating the impending encounter between a standard-sized vehicle and a large-sized vehicle, each traveling in separate lanes. Above: The total risk potential fields generated by the two vehicles traveling at a speed of 25 m/s in different lanes. Below: The vector field of forces generated by the two vehicles at the same speed.

## V. Data Statistics And Validation

For a considerable duration, validating the driving risk potential field model has remained a challenge in this field, typically accomplished through simulation and emulation. Despite effectively reflecting the level of risk for vehicles, practical validation remains elusive. As discussed in the preceding section, it is crucial to note that the validation of risk potential fields can be grounded in the dynamic interactions between vehicles. As a controlled vehicle approaches a target vehicle, its motion state continues to be influenced by neighboring vehicles due to the presence of the target vehicle's risk potential field. Risk force field analysis adeptly elucidates this phenomenon. With this insight, we can concentrate on statistically analyzing the changes in the motion states of vehicles overtaking large-sized vehicles from adjacent lanes. The consistency of the changes in their motion characteristics with the spatial distribution can effectively validate the existence of vehicle risk potential fields.

In this section, we conducted validation and statistical analysis using data from the NGSIM project (FHWA and NGSIM, 2006), specifically from the I-80 freeway section in Emeryville, California, within the San Francisco Bay Area. The comprehensive dataset spans three times intervals on April 13, 2005, from 4:00 PM to 4:15 PM, 5:00 PM to 5:15 PM, and 5:15 PM to 5:30 PM. These intervals represent the escalation of traffic congestion, non-congested periods, transitions between congested and non-congested states, and peak congestion periods, respectively. The data were collected from a six-lane section of the I-80 freeway. The dataset consists of video clips captured by seven synchronized digital video cameras, which were then transcribed into vehicle trajectory data. These trajectory data provide precise vehicle locations within the study area every 0.1 seconds, representing one of the most detailed and accurate microsimulation research data collected to date.

The schematic diagrams of the model we employed are depicted in Fig. 7 and 8. Therefore, this validation specifically involves trajectory data of target vehicles (vehicle type 2 in the dataset) overtaking large vehicles (vehicle type 3 in the dataset) from adjacent lanes. In terms of data processing, we eliminate temporal disarray noise caused by multiple overtakings, retaining only valid data from single successful overtakings. This validation utilizes three data samples spanning a total duration of 45 minutes from the I-80 dataset, collecting all trajectory data that meet the aforementioned criteria. As a result, we collected a sample of 111 large-sized vehicles with 575 standard-sized vehicles that successfully completed overtaking maneuvers. We recorded the trajectory and motion characteristic data of standard-sized vehicles within a range of 12 meters both in front of and behind the large-sized vehicles. As an initial statistic, the average length of large-size vehicles is 13.9 m. Additionally, we categorized large-sized vehicles into 3 groups with lengths ranging from $6-12$ m, $12-18$ m, and $18-24$ m, with average lengths of 8.6 m, 15.9 m, and 20.5 m, respectively.

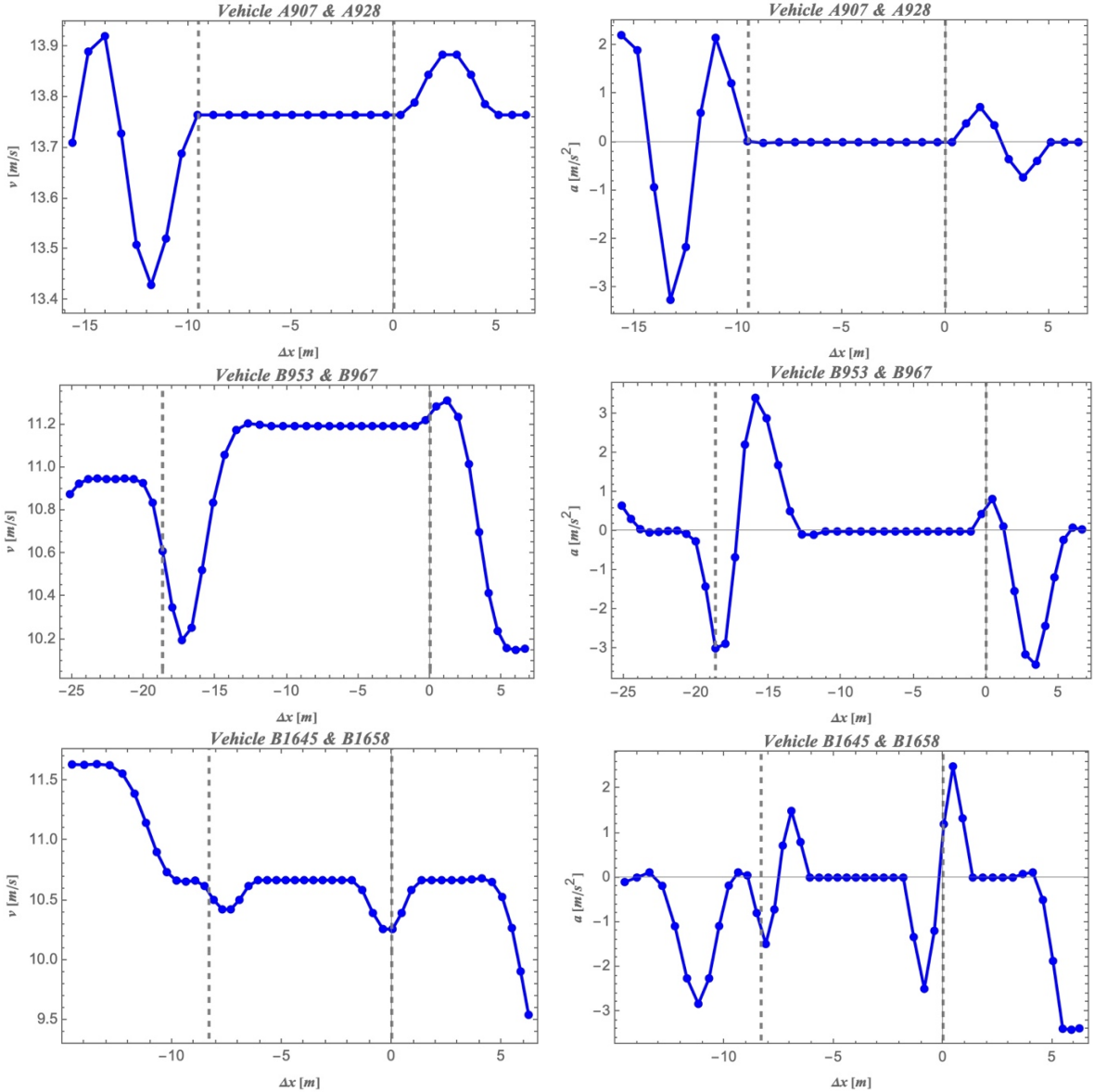

**Fig. 9.** The typical examples of vehicle speed (left) and acceleration (right) as the function of the headway distance between the standard-sized vehicle and large-sized vehicle in the dataset.

Before delving into the discussion, one more point that require clarification. The vehicle accelerations collected in the dataset only reflect numerical changes in the longitudinal direction of the lane. The positive or negative sign represents the direction of acceleration, while the numerical value reflects the intensity of the force field received by the vehicle. Therefore, in the following discussion, we describe increases or decreases in acceleration solely based on numerical changes, without considering the influence of positive or negative signs.

As a demonstration, we selected three typical examples from the dataset. Using the headway distance between the two vehicles at the same moment as coordinates, we plotted the speed and acceleration curves of standard-sized vehicles during the entire overtaking maneuver. As depicted in Fig. 9, we delineated the body length of each large-sized vehicle with vertical gray dashed lines in each example. It is evident that, under relatively ideal conditions, standard-sized vehicles exhibit significant variations in motion state at the tail and head regions of large-sized vehicles, while maintaining a relatively stable motion state at the side region. This is reflected in the acceleration of vehicles at the side region, which remains close to zero over a considerable distance, indicating nearly constant linear motion. However, at the tail and head areas, the scenario is notably different. Taking vehicles B953 and B967 as examples, when vehicle B967 approaches the tail region of the large-sized vehicle B953, it experiences repulsive force from the risk potential field generated by vehicle B953. As a result, vehicle B967 exhibits negative acceleration, causing a rapid decrease in speed. Subsequently, to complete the overtaking maneuver, the driver increases horsepower to accelerate the vehicle, leading to a transition from negative to positive acceleration. Upon leaving the tail region, the vehicle maintains a constant speed alongside vehicle B953. As the vehicle reaches the head region, it experiences a propulsive force from the risk potential field, resulting in a slight increase in acceleration and speed. To counteract the propulsive force, the driver quickly releases the throttle, causing the acceleration to become negative again, further reducing the speed to maintain a relatively stable state as it leaves the head region.

As a demonstration, we selected three typical examples from the dataset. Using the headway distance between the two vehicles at the same moment as coordinates, we plotted the speed and acceleration curves of standard-sized vehicles during the entire overtaking maneuver. As depicted in Fig. 9, we delineated the body length of each large-sized vehicle with vertical gray dashed lines in each example. It is evident that, under relatively ideal conditions, standard-sized vehicles exhibit significant variations in motion state at the tail and head regions of large-sized vehicles, while maintaining a relatively stable motion state at the side region. This is reflected in the acceleration of vehicles at the side region, which remains close to zero over a considerable distance, indicating nearly constant linear motion. However, at the tail and head areas, the scenario is notably different. Taking vehicles B953 and B967 as examples, when vehicle B967 approaches the tail region of the large-sized vehicle B953, it experiences repulsive force from the risk potential field generated by vehicle B953. As a result, vehicle B967 exhibits negative acceleration, causing a rapid decrease in speed. Subsequently, to complete the overtaking maneuver, the driver increases horsepower to accelerate the vehicle, leading to a transition from negative to positive acceleration. Upon leaving the tail region, the vehicle maintains a constant speed alongside vehicle B953. As the vehicle reaches the head region, it experiences a propulsive force from the risk potential field, resulting in a slight increase in acceleration and speed. To counteract the propulsive force, the driver quickly releases the throttle, causing the acceleration to become negative again, further reducing the speed to maintain a relatively stable state as it leaves the head region.

Based on these observations, we conducted a statistical analysis of the trajectories of 575 events of vehicles crossing large-sized vehicles in adjacent lanes. The results, illustrated in Fig. 10, depict the headway distance between the analyzed vehicles and the large-sized vehicles on the horizontal axis, while the vertical axis represents the statistical average speed or acceleration of the vehicles. The average length of large-sized vehicles is denoted in the figure by vertical gray dashed lines indicating the positions of the vehicle's tail and head. By employing high-order polynomial fitting, we produced fitted curves to represent the statistical data, depicted by the blue curves in the figures. These curves adeptly illustrate the motion patterns of analyzed vehicles as they approach and overtake large-sized vehicles in adjacent lanes.

As depicted in the left part of Fig. 10, as vehicles approach the rear area of the large-sized vehicle on the adjacent lane, their average speeds undergo brief oscillations, signifying the onset of influence by the large-sized vehicle. Subsequently, the speeds gradually increase until they depart from the rear area, indicating acceleration to leave the rear region of the large-sized vehicle. Furthermore, during the merging and overtaking of the large-sized vehicle, there is a slight increase until the overtaking is completed, followed by a nearly smooth and stable speed reduction.

Moreover, importantly, the statistical results maintain good consistency with the analysis in Fig. 9. This pattern suggests the presence of a stable risk force field around the large-sized vehicle. As vehicles enter the force field region, they experience a repulsive force from the risk force field, prompting them to counteract this force by accelerating to depart from the tail region. Similarly, when vehicles merge with the large-sized vehicle, they encounter a propulsive force from the risk force field, resulting in an increase in speed. To counteract this propulsive force and prevent the vehicle from losing control, the vehicles reduce their speed. The relationship between vehicle acceleration and distance can further support our previous analysis. As depicted in the right panel of Fig. 10, vehicles undergo significant acceleration in the rear and front regions of the large-sized vehicle due to repulsive force and propulsive force from oblique directions. When vehicles enter the lateral area of the large-sized vehicle, they encounter repulsive forces only perpendicular to the direction of the lane. This leads to oscillations of acceleration around zero, indicating that drivers are continuously adjusting their vehicles to prevent instability.

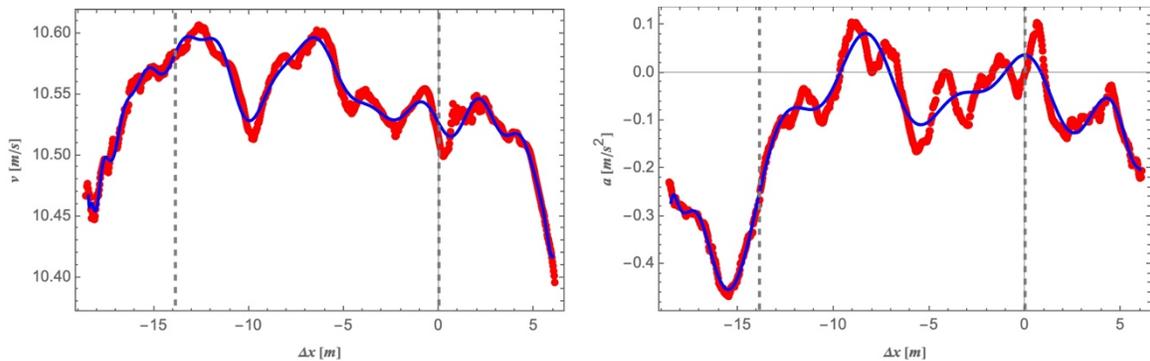

**Fig. 10.** The average speed and average acceleration rate of all target vehicles overtaking large-sized vehicles in the adjacent lane as the function of the headway distance, respectively.

To ensure the comprehensiveness and effectiveness of the statistics, we further analyzed the statistical results under different length ranges of large-sized vehicle bodies. As illustrated in Fig. 11, it is evident that the results of the Group $6 - 12$ m exhibit more regularity, with significant oscillations when vehicles enter the tail area and notable speed advancement when vehicles start overtaking, further validating our earlier analysis. In summary, the results of grouped statistics align with our conclusions: vehicles accelerate to leave the force field area to counteract the repulsive effect of the force field when entering the tail region, and they accelerate due to the propulsive effect of the force field when entering the front region. Subsequently, vehicles decelerate to maintain stability in vehicle movement to counteract the propulsive effect of the force field.

Furthermore, we observe that the changes in acceleration and velocity of the Group $6 - 12$ m vehicles are more pronounced when they are beside the large-sized vehicles. This is because the length of the large-sized vehicle is relatively short at this point, making it easier for vehicles in the adjacent lane to accelerate away from the lateral area. The statistical outcomes of this group display less noise and greater regularity, attributable to the same factor. Conversely, the fluctuations in acceleration statistics for the Group $18 - 24$ m are more intense and oscillatory. Several objective factors contribute to this phenomenon: Firstly, there are fewer large-sized vehicles with extremely long bodies, resulting in a relatively small samples for statistics. Secondly, the risks posed by vehicles with extremely long bodies are greater, and the sense of oppression is stronger, leading to a significantly increased probability of adjacent lane vehicles experiencing instability during travel. Another reason is that our vehicle risk potential field model only considers the ideal situation of interaction between two vehicles, while real vehicle behaviors in the dataset are much more complex. This implies that when we statistically analyze the motion characteristics of vehicles in adjacent lanes to large-sized vehicles, their motion states will inevitably be influenced by surrounding vehicles.

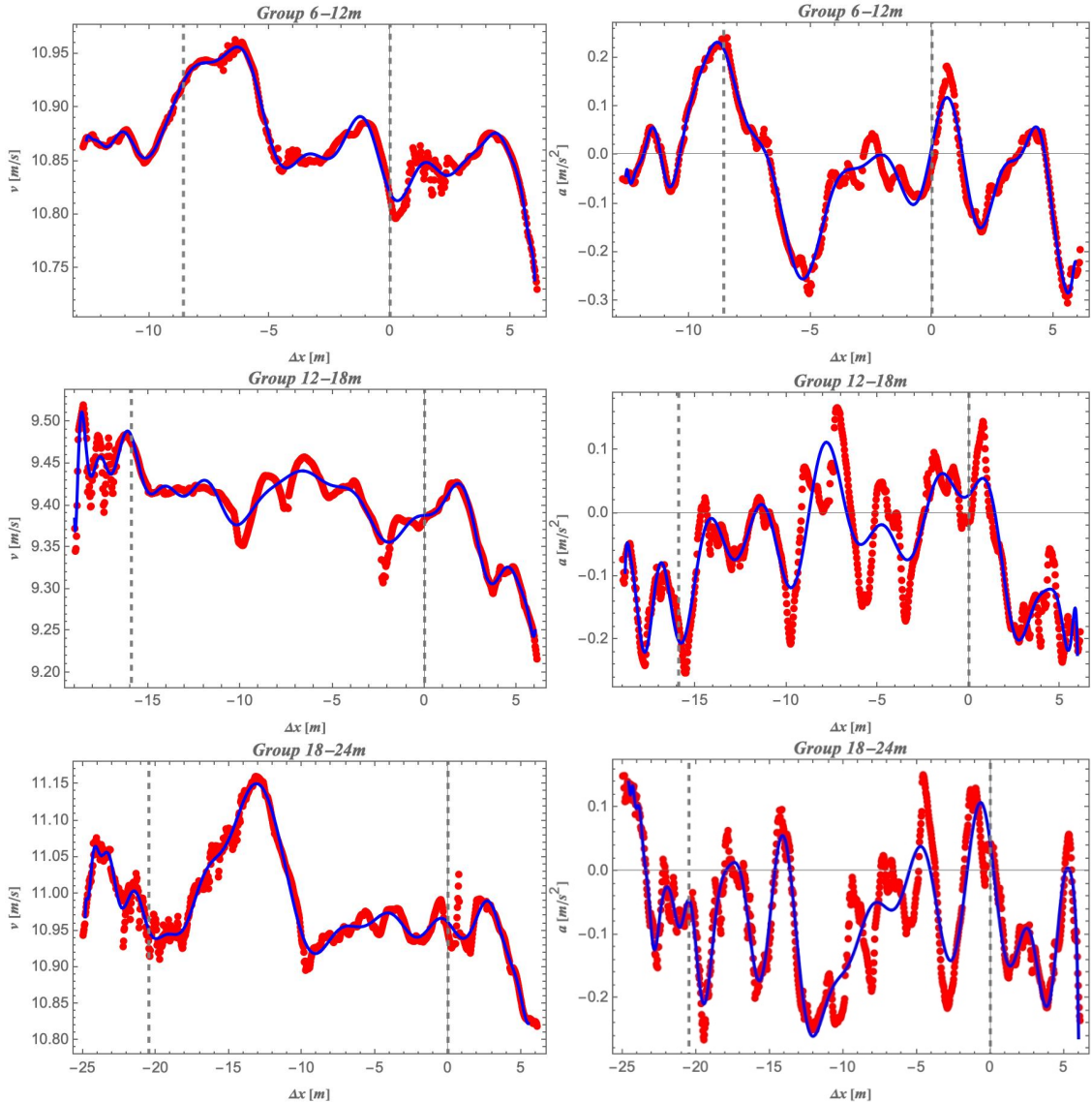

**Fig. 11.** The graph illustrates the mean speed/acceleration of target vehicles overtaking vehicles of different sizes with longitudinal lengths of $6 - 12$ m, $12 - 18$ m, and $18 - 24$ m, from top to bottom, respectively.

Although it is difficult to eliminate this dynamic noise, because it is one of the main factors causing instability in vehicle motion states, according to the inference from ideal results, when vehicles enter the force field range of large-sized vehicles, they continue to be affected by stable force fields, resulting in regular and stable changes in vehicle motion states. Therefore, our statistical results can qualitatively verify the existence of the force field.

## VI. Conclusion And Discussion

In this article, we have constructed a highway risk potential field model that incorporates vehicle dimension factors. By quantitatively simulating the static and dynamic environments of the highway, we have defined road potential fields, lane potential fields, and vehicle potential fields to comprehensively construct a mathematical model for the total risk potential field. An important innovation of this article is the inclusion of simulated corrections for vehicle dimensions in the vehicle potential field. This correction is significant for applying field theory mathematical models to simulate highways. We argue that the impact of vehicle size is not only reflected in the spatial occupation of vehicles during model construction but also profoundly affects the spatial distribution and diffusion of risk potential fields.

Using this approach, we analyze the total risk potential fields formed by standard vehicle, medium-sized vehicle, and large-sized vehicle traveling at a constant speed on highways as shown in Fig. 5. The results show that as vehicle size increases, not only does the risk potential field strengthen in front and behind target vehicle but it also spreads to greater distances from it. Additionally, it leads to large areas with certain intensities of risk potentials forming on their sides. Larger vehicle sizes result in increased intensity around vehicle and larger affected areas within these potentials. This means that when larger-sized vehicles are present on highways while driving, total driving risks increase further.

The conclusion is consistent with our driving experience. When we drive near medium or even large-sized vehicles, we can clearly feel the sense of oppression they generate. This intangible pressure is a manifestation of quantified risk potential field. Furthermore, objectively speaking, factors such as larger blind spots in bigger vehicles, greater vehicle inertia, more complex vehicle handling, and the more intricate airflow structures formed during high-speed driving all imply that the driving risks around larger vehicles are higher. One major highlight of this article is that we have constructed a quantified risk potential field model by introducing vehicle size to incorporate these complex, variable, and difficult-to-quantify influencing factors.

In addition, we have further optimized the spatial shape of the vehicle by replacing the wedge-shaped area mentioned in Ref. [2] with a curved region. This optimization method can also induce controlled vehicles' drivers to make lane-changing decisions and make the vehicle's potential field smoother in the surrounding space. This is one of the important factors that need to be considered when constructing a model because lane-changing decisions are complex decisions that involve multiple factors. Factors such as following distance, own vehicle speed, and target vehicle speed ahead can be evaluated using risk potential fields. However, more importantly, environmental factors such as distribution of vehicles in adjacent lanes, density of vehicles on surrounding roads, and driver's psychological state all influence lane-changing decisions. Therefore, we hope that the spatial distribution behavior of risk potential fields can be more moderate and our mathematical model can comprehensively consider all surrounding vehicle distributions of controlled vehicle so as to simulate lane-changing decision-making behaviors more accurately and logically. On the other hand, we also examined the impact of vehicle speed on risk potential fields. As shown in Fig. 4, increasing vehicle speed intensifies the diffusion of risk potential fields within lanes, requiring longer following distances to ensure driving safety. This is consistent with relevant regulations on highways and further validates the effectiveness and feasibility of our mathematical model in this paper.

Furthermore, another significant innovation of this paper lies in our examination of the interactions between vehicles traveling at high speeds on highways by calculating the force field formed by the driving risk potential field. The concept of a force field is widely used in physics and engineering to describe the influence of various forces present in space on the motion state of objects. When controlled vehicles approach or overtake target vehicles on adjacent lanes, the vector nature of the force field causes the controlled vehicles to experience repulsive and propulsive forces in the tail and head regions, respectively, thereby affecting the motion state of the controlled vehicles. Through the statistical analysis of the motion characteristics of vehicles traveling on adjacent lanes of large-sized vehicles, we have effectively validated the force fields enveloping the vehicles and their corresponding driving risk potential fields. The validation results clearly demonstrate that the model maintains consistency in various driving scenarios and aligns closely with observational data. Additionally, it effectively indicates that, for adjacent lanes, the regions around the tail and head of high-speed vehicles represent highly risky areas.

Unlike traditional traffic models, which study vehicle following and lane-changing behaviors in one dimension, the emergence of the driving risk potential field model naturally lends itself to studying traffic behavior in space and studying interactions among multiple vehicles on complex roads. In this paper, we preliminarily applied the driving risk potential field to study the driving behavior of vehicles on adjacent lanes, effectively validating the model's effectiveness and practical applicability. This encourages us to focus more on the research and analysis of spatial traffic behavior, particularly the modeling and theoretical analysis of lateral traffic behavior. This has significant implications for the analysis of traffic accidents, vehicle-road coordination in complex road environments, and the advancement of future intelligent autonomous driving research, offering considerable research prospects.

## Acknowledgment

The authors thank the helpful discussions with Dr. Hang Liu and Dr. Hong Guo.